# Weak field magnetoresistance of narrow-gap semiconductor InSb

R. Yang,[1] K. H. Gao,[1,2] Y. H. Zhang,[1] P. P. Chen,[1, a)] G. Yu,[1,b)] L. M. Wei,[1] T. Lin,[1] N. Dai[1], and J. H. Chu[1,2]

[1]*National Laboratory for Infrared Physics, Shanghai Institute of Technical Physics, Chinese Academy of Science, Shanghai 200083, People's Republic of China*

[2]*Key Laboratory of Polar Materials and Devices, Ministry of Education, East China Normal University, Shanghai 200062, People's Republic of China*

The weak antilocalization effect of InSb film in perpendicular as well as tilted magnetic field is investigated. It is found that the InSb film has quasi-two-dimensional feature and the Nyquist mechanism dominates decoherence. The two dimensionality is also verified further and the influence of roughness effect and Zeeman effect on weak antilocalization effect is studied by systematically investigating the anisotropy of weak field magnetoresistance with respect to magnetic field. It is also found that the existence of in-plane field can effectively suppress the weak antilocalization effect of InSb film and the roughness effect plays an important role in the anisotropy.

Corresponding Author: *E-mail: a) ppchen@mail.sitp.ac.cn
b) yug@mail.sitp.ac.cn

**I. Introduction**

Weak antilocalization effect arises from quantum interference between backscattered time-reversed trajectories of conduction carrier and results in a deviation (quantum correction) to the classical resistance. With the introduction of a perpendicular magnetic field, the time-reversal symmetry will be destroyed and the quantum correction from weak antilocalization effect will vanish. The resulting shape of magnetoresistance curve can reveal information about phase coherent time and spin-orbit scattering time. Thus weak antilocalization is a useful tool for the analysis of spin-orbit interaction in semiconductors.

Narrow-band semiconductors such as InSb have narrow energy gap and strong spin-orbit interaction. They are a class of promising materials that can be used in spintronics.[1-3] In the research of magnetoresistance, anisotropy is an interesting topic which can reveal information about roughness effect and the Zeeman effect. The anisotropy of weak field magnetoresistance can fall into two categories. The first category is the trivial case in which the magnetoresistance curves in tilted magnetic field collapse into those in perpendicular field. This kind of anisotropy has been widely observed in many systems. In the second category the magnetoresistance curves in tilted magnetic field don't collapse into those in perpendicular field. This non-trivial anisotropy has been identified and systematically studied in quantum well systems,[4-7] MOSFET[8] and a few films.[9-11] Two mechanisms have been proposed as possible explanations for this anisotropy. One is the random Berry phase induced by in-plane field–the roughness effect;[4, 8] the other is a mixing of singlet and triplet

Cooperon due to Zeeman coupling—the Zeeman mechanism.[4-7] There have been quite few researches concerning the anisotropy of magnetoresistance in thin films and the mostly anisotropy investigated is in pure weak localization effect. Some found that weak localization effect in perpendicular field is different from that in parallel field and in-plane field can even change the sign of weak field magnetoresistance. The Zeeman coupling is a possible mechanism in this anisotropy.[10, 11] To our best knowledge, there have been no researches dedicated to non-trivial anisotropy in weak antilocalization effect in narrow-gap semiconductors.

In this paper, we investigate weak antilocalization effect and analyze the data with the 2-dimensional (2D) weak antilocalization models. We also systematically investigate the anisotropy of weak field magnetoresistance. It is found that in-plane magnetic field can effectively suppress the weak antilocalization effect. We find that the Nyquist dephasing mechanism is the dominating dephasing mechanism and the roughness effect plays a key role in the anisotropy of intrinsic InSb film. The Zeeman effect is not important for this case. Our results show that the anisotropy investigated in our InSb sample is different from that discovered in some quantum wells and films, in which the Zeeman effect has a strong influence on anisotropy of weak field magnetoresistance.[4-7, 9-11]

## II. Sample fabrication and measuring system

The InSb film used here was grown on a semi-insulating GaAs <100> substrate in a RIBER 32 R&D molecular beam epitaxy (MBE) facility. The growth process was monitored with a Reflection High-Energy Electron Diffraction (RHEED) system. The growth temperature was 390 ℃. The V-III flux ratio was kept approximately 2:1. The average growth rate was 10 nm/min. The thickness was 1 μm. The conduction type was n-type, with the carrier concentration of ~$10^{22}$ m$^{-3}$ and the mobility of ~1.5 m$^2$V$^{-1}$s$^{-1}$ at 1.3~2.0 K according to the Hall measurement. Indium was soldered onto the InSb surface to facilitate Ohmic contacts. The magnetoresistance is measured under Van der Pauw configuration and the magnetic field is applied perpendicular to the film and then tilted. All measurements are performed in an Oxford Instruments $^4$He cryogenic system with the temperature ranges from 1.3 to 2.1 K. A package of Keithley sourcemeters is used to provide current and to measure magnetoresistance.

### III. Experimental results and discussions

#### 1. In perpendicular magnetic field

Weak localization effect with spin-orbit interaction has been observed in our InSb sample (see Fig. 1). The negative magnetoconductivity before the emerging of positive magnetoconductivity is a indication of strong spin-orbit interaction. Moreover, with the increase of temperature, we can see that the negative magnetoconductivity 'cusp' fades away gradually. This is because the depth of the 'cusp' is proportional to $\tau_\varphi/\tau_{so}$ ($\tau_\varphi$ is the phase coherent time and $\tau_{so}$ is the spin-orbit scattering time), with the increase of temperature, $\tau_\varphi$ decreases and $\tau_{so}$ is basically temperature-independent.[12]

There are two types of models concerning weak localization effect in films. One is the weak localization model in 2D system which is used by many research groups for the weak localization effect in semiconductor quantum wells and films.[12-14] The other type is the weak localization model in 3-dimensional (3D) system.[15-19] The dimension for a system in which the phase coherent phenomena occurs depends on the relative magnitudes of the phase coherent length $L_\varphi = \sqrt{D\tau_\varphi}$ ($D$ is the diffusion constant) and the film thickness $t$. When $L_\varphi \approx t$, the system should be regarded as a 2D system while $L_\varphi \ll t$ is regarded as a 3D system.[12, 20-22]

The weak localization model in 2D regime has various versions. Here, we use the ILP (S. V. Iordanskii, Y. B. Lyanda-Geller, G. E. Pikus) model.[4, 14]

For the ILP model,[4] the weak localization correction to the Drude conductivity

$$\Delta\sigma = \Delta\sigma(B) - \Delta\sigma(0) = \frac{G_0}{2}\left(F_t(b_\varphi, b_s) - F_s(b_\varphi)\right) \quad (1)$$

where $G_0 = e^2/2\pi^2\hbar$; $b_\varphi = B_\varphi/B$; $b_s = B_s/B$; $B_\varphi = \hbar/4eD\tau_\varphi$; $B_s = \hbar/4eD\tau_{so}$; and $e$ is the electronic charge, $\hbar$ is Plank's constant divided by $2\pi$, $D$ is the diffusion constant and it's determined from the Hall measurement and the resistivity at $B=0$. $F_s(b_\varphi, b_s)$ is the triplet contribution, corresponding to scattered electrons who have the total momentum equal to one. $F_t(b_\varphi, b_s)$ is the singlet contribution, corresponding to scattered electrons which have the total momentum of zero. They are given by

$$F_s(b_\varphi) = \psi(\frac{1}{2} + b_\varphi) - \ln b_\varphi \quad (2)$$

$$F_t(b_\varphi, b_s) = \sum_{n=1}^{\infty}\left\{\frac{3}{n} - \frac{3a_n^2 + 2a_n b_s - 1 - 2(2n+1)b_s}{(a_n + b_s)a_{n-1}a_{n+1} - 2b_s\left[(2n+1)a_n - 1\right]}\right\} - \frac{1}{a_0} - \frac{2a_0 + 1 + b_s}{a_1(a_0 + b_s) - 2b_s}$$
$$-2\ln(b_\varphi + b_s) - \ln(b_\varphi + 2b_s) - 3C - S(b_\varphi/b_s) \quad (3)$$

Here $S(x)$ and $a_n$ are given by

$$S(x) = \frac{8}{\sqrt{7+16x}}\left[\arctan(\frac{\sqrt{7+16x}}{1-2x}) - \pi\,\Theta(1-2x)\right] \quad (4)$$

$$a_n = n + \frac{1}{2} + b_\varphi + b_s \quad (5)$$

where $\psi(x)$ is the digamma function, $C$ is the Euler constant, $\Theta(x)$ is the Heaviside step function.

The curve fitting results are shown in Fig. 1. All fitted curves show that $\tau_\varphi$, $\tau_{so}$ consistently agree with previous research results reported[9, 12] in magnitude ($\tau_\varphi$ ranges from $10^{-12}$–$10^{-10}$ s and $\tau_{so} \approx 8.7 \times 10^{-12}$ s). And $\tau_\varphi$ is self-consistently found to give rise to $L_\varphi$ (0.5 μm), comparable to $t$ (1 μm). This justifies the application of 2D models. In contrast, fitting the data with 3D model (like revised Kawabata model[15]) gives rise to $L_\varphi$ (0.7 μm) ~$t$ (1 μm), contradicting with the prerequisite of 3D model ($L_\varphi \ll t$).[15, 23] This concludes that the relatively large $L_\varphi$ makes our InSb film having an obvious 2D nature.

From the relation between $\tau_\varphi$ and $T$, we can get the information about the decoherence mechanism in InSb thin film. Generally speaking, $\tau_\varphi \sim T^p$, and p=1 or p=3 are corresponding to the Nyquist decoherence mechanism[12] or the electron-phonon interaction decoherence mechanism, respectively.[24, 25] For our data, the best fit is obtained for p=1, corresponding to the Nyquist decoherence mechanism as shown in Fig.2. The result agrees with those reported in a recent publication.[12]

The Nyquist decoherence mechanism[12] predicts a linear relation between $\tau_\varphi$ and $T^{-1}$:

$$\tau_\varphi = \alpha \bullet \frac{1}{T}, \alpha = \frac{\frac{2\pi\hbar^2}{e^2}\sigma_0 t}{k_B \ln\left(\frac{\pi\hbar}{e^2}\sigma_0 t\right)} \tag{6}$$

where $\alpha$ is the dephasing time, $\sigma_0$ is the zero-field conductivity, and $t$ is the thickness of the film.

With $\tau_\varphi$ obtained from the ILP model, after fitting $\tau_\varphi \sim T^{-1}$ curve with linear equation, (see Fig.2), we have a dephasing time ($0.3 \times 10^{-10}$ s) different the theory value ($1.87 \times 10^{-10}$ s) obtained from equation (6). The discrepancy between the experimental value and theoretical value is acceptable, since it is within the measurement uncertainty. Such a discrepancy also exists in previous research results for ranges from a half to orders of magnitude.[12, 26]

$\tau_{so}$ is basically temperature-independent ($\tau_{so} \approx 8.7 \times 10^{-12}$ s). This result is quite reasonable because there are two kinds of spin-orbit scattering mechanisms: the D'yakonov-Perel (DP) mechanism and the Elliott-Yafet (EY) mechanism, determining spin decoherence in n-type InSb. Both of them can lead to temperature-independent $\tau_{so}$.

According to $\tau_{so}$ obtained from above analysis, we can evaluate $L_{so} \approx 0.4\ \mu m$ ($L_{so}$ is the spin coherent length), close to value(~1 $\mu m$) reported in previous study.[12] This is a considerably long length and can be readily achieved with modern techniques, allowing fabrication of devices based on spin-coherent phenomena.

**2. In tilted magnetic field**

Investigating magnetoresistance in tilted field can reveal much information about the dimension and the anisotropy of magnetoresistance. The anisotropy of weak antilocalization effect has been predicted[27] and observed[4-7] in quantum wells. The anisotropy has been systematically investigated in InGaAs and GaAlN quantum wells.[4-7] To our best knowledge, it has not been studied in details in narrow-band semiconductor thin film systems. Here, we investigate this phenomenon in details.

We can see from Fig. 3(a) that the weak antilocalization effect (the cusp-like negative magnetoconductivity part of the magnetoconductivity curve) is suppressed in tilted magnetic field. The magnetoconductivity curve doesn't collapse into the curve corresponding to the case in which the perpendicular-field is applied when $B$ is replaced by $B_\perp$, thus the anisotropy is nontrivial. This effect has been widely observed in previous experiments concerning weak field magnetoresistance of quantum well systems.[4-7]

The anisotropy is a strong evidence of 2D nature of the system. The basic physics behind this phenomenon is that the in-plane field can lead to the suppression of weak antilocalization effect through the Zeeman effect and roughness effect.[4, 6] We fit our data with the models proposed in Ref. 4 that investigated the suppression of weak antilocalization of InGaAs quantum well in a tilted magnetic field.

According to Ref. 4, the magnetoconductivity of 2DEG in a tilted field can be expressed as:

$$\Delta\sigma = \Delta\sigma(B) - \Delta\sigma(0) = f(B_\perp, B_\parallel; g, d^2L, \tau_\varphi, \tau_{so}) = \frac{G_0}{2}\left(F_t(b_\varphi, b_s) - F_s(b_\varphi)\right) \quad (7)$$

where $G_0 = \dfrac{e^2}{2\pi^2\hbar}$; $b_\varphi = \dfrac{B_\varphi + \Delta_r(B_\parallel)}{B_\perp}$; $b_s = \dfrac{B_s}{B_\perp}$; $b_\varphi = \dfrac{B_\varphi + \Delta_r(B_\parallel) + \Delta_s(B_{tot})}{B_\perp}$;

$B_{tot} = \sqrt{B_\parallel^2 + B_\perp^2}$

$\Delta_r(B_\parallel) \simeq \dfrac{\sqrt{\pi}}{2}\dfrac{e}{\hbar}\dfrac{d^2L}{l}B_\parallel^2$; $\Delta_s(B_{tot}) = \dfrac{\tau_{so}}{4e\hbar D}(g\mu_B B_{tot})^2$;

$B_\varphi = \dfrac{\hbar}{4eD\tau_\varphi}$; $B_s = \dfrac{\hbar}{4eD\tau_{so}}$

$\mu_B$ is Bohr magneton, $g$ is the Lande factor, $d^2L$ is a value that measures the roughness, $l = v_F\tau_p$ is the mean free path, and $v_F$ is the Fermi velocity. $\tau_p$, $D$ are momentum scattering time and diffusion constant, respectively. They are determined from the Hall measurement and the resistivity at $B=0$.

$F_s(b_\varphi)$ is the triplet contribution, corresponding to scattered electrons who have the total momentum equal to one. $F_t(b_\varphi, b_s)$ is the singlet contribution, corresponding to scattered electrons which have the total momentum of zero. They are given by

$$F_s(b_\varphi) = \psi(\frac{1}{2}+b_\varphi) - \ln b_\varphi \tag{8}$$

$$F_t(b_\varphi, b_s) = \sum_{n=1}^{\infty}\left\{\frac{3}{n} - \frac{3a_n^2 + 2a_n b_s - 1 - 2(2n+1)b_s}{(a_n+b_s)a_{n-1}a_{n+1} - 2b_s\left[(2n+1)a_n - 1\right]}\right\} - \frac{1}{a_0} - \frac{2a_0 + 1 + b_s}{a_1(a_0+b_s) - 2b_s}$$
$$-2\ln(b_\varphi + b_s) - \ln(b_\varphi + 2b_s) - 3C - S(b_\varphi / b_s) \tag{9}$$

Here $S(x)$ and $a_n$ are given by

$$S(x) = \frac{8}{\sqrt{7+16x}}\left[\arctan(\frac{\sqrt{7+16x}}{1-2x}) - \pi\,\Theta(1-2x)\right] \tag{10}$$

$$a_n = n + \frac{1}{2} + b_\varphi + b_s; \tag{11}$$

where $\psi(x)$ is the digamma function, $C$ is the Euler constant, $\Theta(x)$ is the Heaviside step function.

Generally speaking, fitting data with Eq. (7) is a surface fitting problem. In our experiment, $B_{//} = B_\perp \tan\theta$, $\theta$ is the angle between the direction of magnetic field and the direction perpendicular to the film, so it can be simplified into a curve fitting problem.

$$\Delta\sigma = \Delta\sigma(B_\perp, B_\parallel) - \Delta\sigma(0, B_\parallel) = f'(B_\perp; g, d^2L, \tau_\varphi, \tau_{so}) \tag{12}$$

The fitting results corresponding to Fig. 3(b) are for $g=-51$; $d^2L=3.8\times10^{-21}$ m$^{-3}$ (the fluctuation volume is $\sim(100\text{ nm})^3$); $\tau_\varphi = 2.4\times10^{-11}$ s; $\tau_{so} = 9.7\times10^{-12}$ s. $\tau_\varphi$, $\tau_{so}$ are close to the values corresponding to perpendicular-field configuration ($2.1\times10^{-11}$ s, $7.9\times10^{-12}$ s respectively), and the little discrepancy between $\tau_\varphi$, $\tau_{so}$ obtained in tilted magnetic field and those in perpendicular field may be due to that the film is not an exact two-dimensional system though it has quasi-two-dimensionality.

Further investigation shows that the anisotropy of magnetoresistance is not sensitive to the Zeeman effect and the roughness effect is more important if we neglect the $\Delta_s$ term or fix g at a slightly different value in equation (7). The fitting result is still good and shows the same $d^2L$, $\tau_\varphi$, $\tau_{so}$. However, if we neglect the roughness term in equation (7), the fitting result is poor. This is different from researches concerning anisotropic magnetoresistance in quantum wells, where the Zeeman effect plays an important role.

According to Mathur and Baranger's work,[8] the mechanism for the influence of roughness on magnetoresistance is due to microroughness. The in-plane uniform magnetic field is then equivalent to a random perpendicular magnetic field due to the microroughness. This equivalent perpendicular magnetic field results in a random Berry phase in electrons when they are bounced around in closed paths and contribute to the vanish of weak antilocalization by increasing the dephasing rate. In our InSb film, the roughness should come from the misfit of the interface between InSb film and GaAs substrate and the film surface fluctuation.

Aside from the roughness effect, in-plane field can also influence the magnetoresistance through the Zeeman coupling. The Zeeman coupling can increase dephasing rate in the singlet contribution of equation (7). The strength of this effect depends on the total magnitude of magnetic field. Thus this effect is only noticeable when the applied field is strong enough.[4, 27] In Ref. 4, the influence of the roughness effect and the Zeeman effect are incorporated in $\Delta_r$ and $\Delta_s$, shown in equation (7). For $\Delta_r$ and $\Delta_s$, which one is dominant depends on its relative magnitude when compared with $B_\varphi$. There are three regimes: the first regime is that $B_\parallel$ is very small and neither $\Delta_r$ nor $\Delta_s$ gains importance; the second regime is that $B_\parallel$ is large enough to make the larger one in $\Delta_r$ and $\Delta_s$ gains prominence; the third regime is that $B_\parallel$ is quite large to make both $\Delta_r$ and $\Delta_s$ come into significance.

In our experiment, $B_\varphi \approx 0.5$ mT, $\Delta_{r\max}$ ~0.1 mT, $\Delta_{s\max}$ ~0.001 mT, $\Delta_s$ is negligible thus the Zeeman effect doesn't come into significance and our sample is in the second regime as mentioned above. As a comparison, Minkov group's research concerns anisotropic magnetoresistance in InGaAs quantum well system,[4] for $B_\varphi \approx 0.2$ mT, $\Delta_{r\max}$ ~0.2 mT, $\Delta_{s\max}$ ~0.2 mT. Both $\Delta_r$ and $\Delta_s$ are comparable with $B_\varphi$ thus both the roughness effect and the Zeeman effect influence the anisotropy of magnetoresistance and their experiment is in the third regime as mentioned above. This is because the quite strong magnetic field they applied ($B_{tot} \approx B_\parallel \sim 0.5T$) and the intrinsic property of InGaAs material that makes $\Delta_r$ and $\Delta_s$ have approximately the same magnitude. As a contrast, in our experiment, $B_{tot}$ ~1 mT, thus the Zeeman effect is completely ignorable. Obviously, in order to make the Zeeman effect having noticeable influence on the anisotropy of magnetoresistance in our InSb sample, we have to increase the in-plane magnetic field while keeping the magnetoresistance curve in the weak localization regime. More specifically, in order to make the Zeeman effect come into significance when the weak localization effect happens in our sample, we have to put the in-plane magnetic field at the magnitude of ~0.1T that renders a Zeeman term comparable to $B_\varphi$ when the magnetoresistance curve is at the weak localization region ($B_\perp$ is kept at the magnitude of mT). This is impossible for our experimental setup. Because in our setup $B_\|$ and $B_\perp$ are at the same magnitude due to $B_\| = B_\perp \tan\theta$. In order to put the in-plane magnetic field at the magnitude (~0.1T) that renders a Zeeman term comparable to $B_\varphi$ when the magnetoresistance curve is at the weak localization region, we have to control $B_\perp$ and $B_\|$ independently, just as some researches concerning anisotropy of quantum-wells do.[4] The potential of investigating the Zeeman effect in the anisotropy of magnetoresistance is the main advantage of the setup that can control $B_\perp$ and $B_\|$ separately.

We notice that the importance of roughness is also qualitatively explained in Ishida group's research results.[9] This observation is consistent with our results. We also notice that the similar anisotropy has also been studied by Nitta group.[5,7] They use a model that only considers the Zeeman effect and conclude that the anisotropy is caused by the time-reversal symmetry broken induced by the Zeeman coupling. Fitting our data with this model gives poor results. This also implies that the model from Ref. 4 is much better and the roughness effect plays an important role when it comes to anisotropy of our sample. In terms of the relative influence of the Zeeman effect and roughness effect, we can say that in the research of Ref. 4, both the Zeeman effect and roughness effect come into significance. In the experiment of Ref. 5, the Zeeman effect dominates the anisotropy; and in our experiment, the roughness effect dominates the anisotropy, thus the relative importance of the Zeeman effect and roughness effect influence the details of anisotropy.

## Ⅳ. Conclusions

We investigated the weak antilocalization effect in InSb thin films and found quasi-two-dimensional features in our samples. By fitting the data with the ILP model, we conclude that the Nyquist mechanism is responsible for the decoherence in our samples. We also observed that the in-plane field can effectively suppress the weak antilocalization effect in InSb films, which is attributed to roughness effect.

**Acknowledgements**

This work was supported in part by the Special Funds for Major State Basic Research under Project No. 2007CB924901, National Natural Science Foundation of China under Grant Nos. 60876059, 60976093 and 60906045, Science and Technology Commission of Shanghai under Grant No. 09JC1415700, and the Open Project of State Key Laboratory of Functional Materials for Informatics.

**References**

[1]B. Datta and S. Das, Appl. Phys. Lett. **56**, 665 (1990).

[2]B. Andrei Bernevig, Taylor L. Hughes, Shou-Cheng Zhang, Science **314**, 1757(2006).

[3]Y. L. Chen, J. G. Analytis, J.-H. Chu, Z. K. Liu, S.-K. Mo, X. L. Qi, H. J. Zhang, D. H. Lu, X. Dai, Z. Fang, S. C. Zhang, I. R. Fisher, Z. Hussain, Z.-X. Shen, Science **325**,178(2009).

[4]G. M. Minkov, A. V. Germanenko, O. E. Rut, and A. A. Sherstobitov, Phy. Rev. B **70**, 155323 (2004).

[5]F. E. Meijer, A. F. Morpurgo, T. M. Klapwijk, T. Koga, and J. Nitta, Phy. Rev. B **70**, 201307(R) (2004).

[6]S. Cabañas, Th. Schäpers, N. Thillosen, N. Kaluza, V. A. Guzenko, and H. Hardtdegen, Phy. Rev. B **75**, 195329 (2007).

[7]F. E. Meijer, A. F. Morpurgo, T. M. Klapwijk, and J. Nitta, Phy. Rev. Lett. **94**, 186805(2005).

[8]H. Mathur and Harold U. Baranger, Phys. Rev. B **64**, 235325 (2001).

[9]Shuichi. Ishida, Keiki. Takeda, Atsushi Okamoto, Ichiro Shibasaki, Physica. E **20**, 211 (2004).

[10]Eric D. Black and John C. Price, Phys. Rev. B **58**, 7844 (1998).

[11]D. Kowal, M. Ben-Chorin, and Z. Ovadyahu, Phys. Rev. B **44**, 9080 (1991).

[12]R. L. Kallaher, J. J. Heremans, Phy. Rev. B **79**, 075322 (2009).

[13]S. Hikami, A. Larkin, and Y. Nagaoka, Prog. Theor. Phys. **63**, 707 (1980).

[14]S. V. Iordanskii, Y. B. Lyanda-Geller, and G. E. Pikus, JETP Lett. **60**, 206 (1994).

[15]M. Oszwaldowski, T. Berus, Phy. Rev. B **65**, 235418 (2002).

[16]R. C. Dynes, T. H. Geballe, G. W. Hull, Jr., and J. P. Garno, Phy. Rev. B **27**, 5188 (1983).

[17]A. Kawabata, Solid State Commun. **34**, 431 (1980).

[18]H. Fukuyama, K. Hoshino, J. Phys. Soc. Jpn. **50**, 2131 (1981).

[19]R. G. Mani, L. Ghenim, and J. B. Choi, Solid State Commun. **79**, 8 (1991).

[20]G. Bergmann, Phys. Rep. **107**, 1(1984).

[21]S. Kobayashi and F. Komori, Prog. Theor. Phys. Suppl. **84**, 224(1985).

[22]Z. Ovadyahu and Y. Imry, Phys. Rev. B **24**, 7439 (1981).

[23]P. A. Lee and T. V. Ramakrishnan, Rev. Mod. Phys. **57**, 287 (1985).

[24]J. Rammer and A. Schmid, Phy. Rev. B **34**, 1352 (1986).

[25]M. Yu. Reizer, Phy. Rev. B **40**, 5411 (1989).

[26]S. A. Studenikin, P. T. Coleridge, N. Ahmed, P. J. Poole, and A. Sachrajda, Phy. Rev. B **68**, 035317 (2003).

[27]A. G. Mal'shukov, K. A. Chao, and M. Willander, Phy. Rev. B **56**, 6436 (1997).

**Figure captions**

Figure 1. Comparisons of curve fitting results for different temperatures. All curves are vertically shifted for clarity. Blue circles are experimental data. Red curves are the fitting result using ILP model.

Figure 2. Plots of $\tau_\varphi$ vs 1/T. Blue triangles are experimental data. The red line is a result of linear fitting.

Figure 3. (a) Blue circles correspond to data obtained from tilted-field setup (the angle θ between the direction of magnetic field and the direction perpendicular to film is 30°) at 1.5K. Green triangles are data obtained from perpendicular-field setup (the angle θ between the direction of magnetic field and the direction perpendicular to film is 0°) at 1.5K. The red curve is a fitting result with ILP model (equation (3)). (b) Blue circles correspond to data obtained from tilted magnetic field setup with θ=30°. Green triangles are data obtained from perpendicular-field setup (the angle θ between the direction of magnetic field and the direction perpendicular to film is 0°) at 1.5K. The red curve is a fitting result with equation (7).

Figure 1

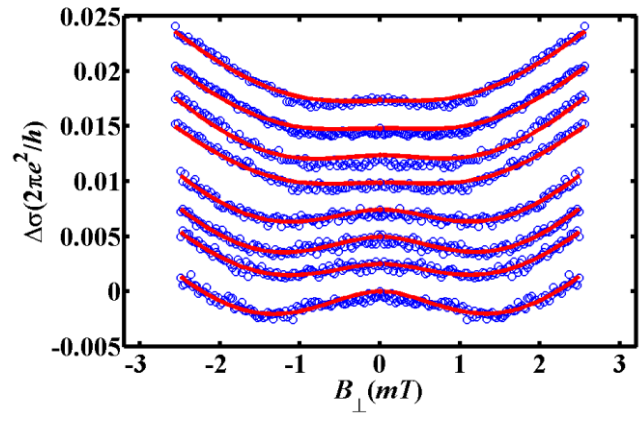

Figure 2

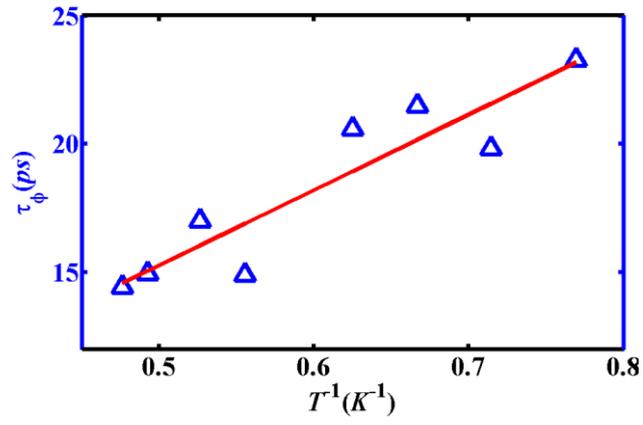

Figure 3

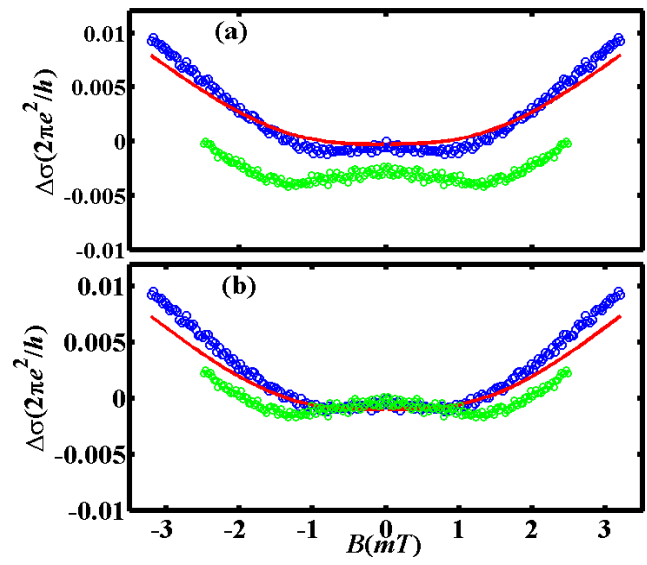